\documentclass{aastex631}

\begin{document}

\title{MeerKAT observations of Procyon at 815.5 MHz}

\correspondingauthor{Ian Heywood}
\email{ian.heywood@physics.ox.ac.uk}

\author[0000-0001-6864-5057]{Ian Heywood}
\affiliation{Astrophysics, Department of Physics, University of Oxford, Keble Road, Oxford, OX1 3RH, UK}
\affiliation{Department of Physics and Electronics, Rhodes University, PO Box 94, Makhanda, 6140, South Africa}
\affiliation{South African Radio Astronomy Observatory, 2 Fir Street, Observatory 7925, South Africa}

\author[0000-0003-2828-7720]{Andrew P.~V.~Siemion}
\affiliation{Breakthrough Listen, University of California, Berkeley, 501 Campbell Hall 3411, Berkeley, CA 94720, USA}
\affiliation{Astrophysics, Department of Physics, University of Oxford, Keble Road, Oxford, OX1 3RH, UK}
\affiliation{SETI Institute, 339 Bernardo Ave, Suite 200, Mountain View, CA 94043, USA}
\affiliation{Department of Physics and Astronomy, University of Manchester, UK}
\affiliation{University of Malta, Institute of Space Sciences and Astronomy, Msida, MSD2080, Malta}

\author{Daniel Czech}
\affiliation{Astrophysics, Department of Physics, University of Oxford, Keble Road, Oxford, OX1 3RH, UK}
\affiliation{Breakthrough Listen, University of California, Berkeley, 501 Campbell Hall 3411, Berkeley, CA 94720, USA}

\author{Steve Ertel}
\affiliation{Steward Observatory, University of Arizona, Tucson, AZ 85721, USA}

\author{Jamie Drew}
\affiliation{The Breakthrough Initiatives, NASA Research Park, Bld. 18, Moffett Field, CA 94035, USA}

\author{Kyran Grattan}
\affiliation{The Breakthrough Initiatives, NASA Research Park, Bld. 18, Moffett Field, CA 94035, USA}

\author{Kevin Wagner}
\affiliation{Steward Observatory, University of Arizona, Tucson, AZ 85721, USA}

\author{S.~Pete Worden}
\affiliation{The Breakthrough Initiatives, NASA Research Park, Bld. 18, Moffett Field, CA 94035, USA}



\begin{abstract}

We have conducted observations of the nearby (11.46 ly) star system Procyon, using MeerKAT's UHF (544--1087 MHz) receivers.
We produce full-Stokes time and frequency integrated continuum images, as well as total intensity time series imaging at 8 s cadence, and full-Stokes vector-averaged dynamic spectra from the visibilities in order to search for transient activity such as flaring events. We detect no significant radio emission from the system, and estimate an upper limit on the circular polarisation fraction of 65 per cent (3$\sigma$ confidence level). A comparison with previous VLA observations places a 3$\sigma$ lower limit on the spectral index between 815.5 and 8400 MHz of 0.26, however long-term significant variability over the last 33 years cannot be ruled out without further, regular radio monitoring of the system.

\end{abstract}



\section{Introduction} 
\label{sec:intro}

Procyon ($\alpha$ Canis Minoris), one of the nearest and brightest stars in the night sky, is part of a 40.84 year period binary system, consisting of a main-sequence F5 IV-V star (Procyon A) and a white dwarf companion (Procyon B) with masses of 1.478 and 0.592~M$_{\odot}$ respectively \citep{bond2015}. Despite its close proximity at just 11.46 ly (3.51 pc) distance, Procyon has not been extensively studied in the radio regime \citep[to the best of our knowledge, the only study to-date being conducted by][]{drake1993}, especially compared to other nearby stars \citep[e.g.][]{triglio2018}. The sensitivity of modern radio instrumentation has led to observations with such facilities becoming increasingly valuable for stellar astrophysics, probing phenomena such as  magnetic field configurations \citep[e.g.][]{bastian2022}, atmospheric temperature measurements \citep{villadsen2014}, coronal mass ejections \citep{crosley2018}, and possibly star-planet interactions \citep{pineda2023}. Furthermore, there are now concerted systematic efforts well underway to observe large numbers of star systems with radio telescopes in pursuit of technosignature detections (e.g. Czech et al.,~\emph{in prep.}). Nearby stellar systems are of course particularly interesting targets in the search for extraterrestrial intelligence, with the closest known earth-like exoplanet being at essentially the same distance as Procyon \citep{bonfils2018}.

\citet{drake1993} observed Procyon at 8.4~GHz with the Karl G.~Jansky Very Large Array over five epochs spanning a 44 day period between April and May of 1991. They detected a moderately varying source with an average intensity of 33~$\mu$Jy beam$^{-1}$ in four of them. For the first of the five epochs the compact radio source had a measured intensity of 115~$\mu$Jy beam$^{-1}$, which is interpreted as either a flare event, or an active region that is subsequently rotated out of view. In this research note we present the (null) results of our investigation of the radio properties of the Procyon system using the MeerKAT radio telescope at a central frequency of 815.5~MHz.

\section{Observations and data processing}
\label{sec:obs}

Procyon was observed\footnote{Project code DDT-20231201-IH-01} using MeerKAT's UHF band (544--1087 MHz) receiver system on 10 December 2023. MeerKAT \citep{jonas2016} is a 64 $\times$ 13.5 m diameter dish radio interferometer array in South Africa's Northern Cape province. It has a densely packed core (70 per cent of the collecting area with a 1~km radius), and a maximum baseline length of approximately 8~km. The primary delay and bandpass calibrator was J0408$-$6545, with J0730$-$1141 used as a secondary gain calibrator, and 3C~138 being used as the polarisation calibrator. Procyon was observed for a total of 4 h 50 m. The {\sc casa} package \citep{casa2022} was used for calibration, and basic flagging of the data, with automated flagging making use of the {\sc tricolour} software \citep{hugo2022}. Imaging and self-calibration were performed with {\sc wsclean} \citep{offringa2014} and {\sc cubical} \citep{kenyon2018} respectively. Total intensity calibration and imaging was automated using the {\sc oxkat} scripts \citep{heywood2020}, whereas polarimetric calibration and imaging was performed manually using the aforementioned tools, with no self-calibration taking place for the polarised imaging.

In addition to full-Stokes, time-integrated multifrequency synthesis images, we produced time series imaging with an 8 s cadence to search for flares or burst-like behaviour from the Procyon system. This was achieved by subtracting a model of the sky (excluding Procyon) from the visibility data and then producing dirty images of the residuals. The same data was then also used to produce full Stokes dynamic spectra of the system by plotting the vector-averaged complex visibilities, averaged over all baselines for every time/frequency interval.

\section{Results and Discussion}
\label{sec:results}

\begin{figure*}
\centering
\includegraphics[width=\textwidth]{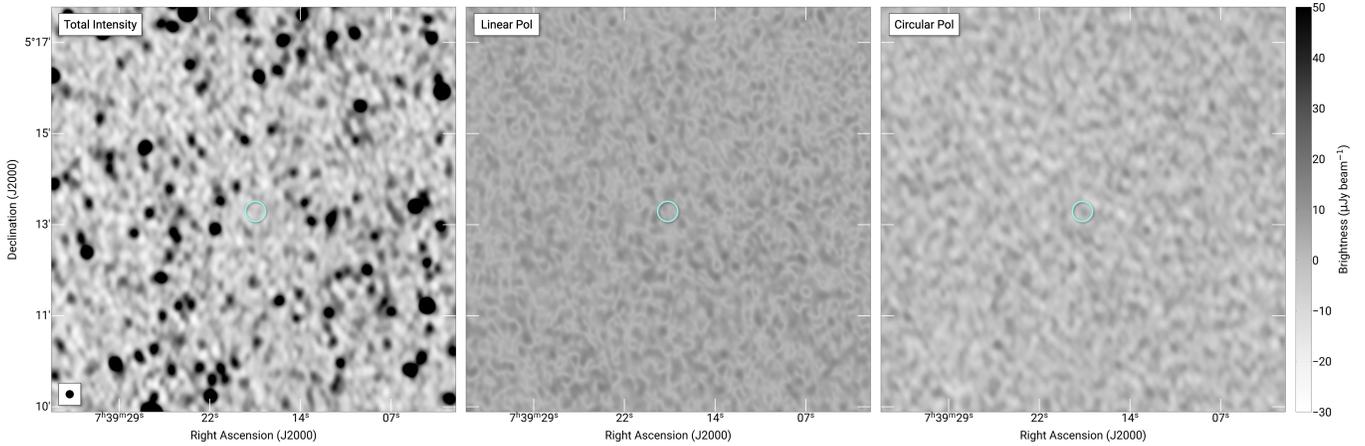}
\caption{The radio brightness of an 8$'$~$\times$~8$'$ region of sky centred on Procyon in (left to right) total intensity (Stokes I), linear polarised intensity ($\sqrt{Q^{2}+U^{2}}$) and circularly polarised intensity (Stokes V). The circle in the lower left corner of the figure shows the circular 9.6$''$ restoring beam used following deconvolution of the total intensity image. The location of Procyon is shown by the central circle. The numerous radio sources visible in the total intensity image are likely predominantly extragalactic.}
\label{fig:procyon}
\end{figure*}

We detect no significant radio emission at the position of the Procyon system in any of the data products described in Section \ref{sec:obs} (see Figure \ref{fig:procyon}). The (1$\sigma$) root mean square noise measurements at the position of Procyon are [6.0,~3.9,~4.3,~3.9] $\mu$Jy~beam$^{-1}$ in the Stokes [I,~Q,~U,~V] maps respectively. The noise in the Stokes I image is elevated by the confusion background, whereas the noise in the Q and U maps is elevated by sidelobe confusion from off-axis sources. The Stokes V image is essentially thermal noise limited. We can put no constraints on the linear polarisation fraction of the system, however we estimate a 3$\sigma$ upper limit of 65 per cent on the circular polarisation fraction. The Stokes I map in combination with the 33~$\mu$Jy~beam$^{-1}$ measurement of \citet{drake1993} places a 3$\sigma$ lower limit on the spectral index\footnote{We adopt the spectral index convention $S$~$\propto$~$\nu^{\alpha}$.} of 0.26, consistent with a thermal emission radio spectrum. Note that the angular resolution of our data (10$''$) is not sufficient to resolve Procyon A from B, the two stars having a maximum angular separation
of 5$''$.

We note that the spectral index limit could easily be biased due to any significant variability in the radio emission from Procyon that has occurred over the 33 years between the two observations used. The 5.1~AU separation at periastron is thought to be sufficient to induce tidal interactions between Procyon A and B \citep{bond2015}, however the temporal separation between the two radio observations covers 80 per cent of the orbital period of the system. The elevated brightness seen in one epoch by \citet{drake1993} suggests that the system exhibits brightness changes by a factor of $\sim$3.5 over timescales of $\sim$2 weeks. Regular multifrequency radio monitoring of Procyon would be required to disentangle these considerations.


\section*{Acknowledgements}

We are grateful to Fernando Camilo for the director's discretionary time award that facilitated this investigation. The MeerKAT telescope is operated by the South African Radio Astronomy Observatory, which is a facility of the National Research Foundation, an agency of the Department of Science and Innovation. Breakthrough Listen is managed by the Breakthrough Initiatives, sponsored by the Breakthrough Prize Foundation. IH thanks Benjamin Hugo for many useful discussions regarding the polarimetric calibration of MeerKAT data.

\bibliographystyle{aasjournal}
\bibliography{procyon}






\end{document}